\documentstyle [12pt,epsf]{article}

\input{epsf}
\begin{document}
\begin{titlepage}
\begin{center}
  
 \vspace{-0.7in}

{\large \bf The Strong-Coupling Expansion and the Ultra-local\\
 Approximation in Field Theory}\\
 \vspace{.3in}{\large\em N.~F.~Svaiter 
\footnotemark[1]}\\
 Centro Brasileiro de Pesquisas Fisicas,\\
 Rua Dr. Xavier Sigaud 150,\\
 22290-180, Rio de Janeiro, RJ Brazil \\
 
\subsection*{\\Abstract}
\end{center}

We discuss the strong-coupling expansion in Euclidean field 
theory. In a formal 
representation for the Schwinger functional, we treat the 
off-diagonal terms of the 
Gaussian factor as a perturbation about the remaining terms 
of the functional integral. 
In this way, we develop a perturbative expansion 
around the ultra-local model, where 
fields defined at different points of Euclidean space are decoupled.  
We first study the strong-coupling expansion in the 
$(\lambda\varphi^{4})_{d}$
theory and also quantum electrodynamics.  Assuming the ultra-local approximation,
we examine the singularities of these perturbative expansions, 
analysing the analytic structure of the zero-dimensional 
generating functions in the complex coupling constants plane.
Second, we discuss the ultra-local generating functional in  
two idealized field theory models
defined by the following interaction Lagrangians: ${\cal{L}_{II}}
(g_{1},g_{2};\varphi)=g_{1}\varphi^{\,p}(x)+g_{2}\varphi^{-p}(x)$, and 
the sinh-Gordon model, i.e., ${\cal{L}_{III}}(g_{3},g_{4};\,\varphi)=
g_{3}\left(\cosh (g_{4}\,\varphi(x))-1\right)$. 
To control the divergences of the strong-coupling perturbative 
expansion two different steps are used throughout the paper. First, 
we introduce a lattice structure to give meaning to the 
ultra-local generating functional. Using an analytic regularization procedure we discuss briefly how it is possible to obtain a renormalized 
Schwinger functional associated with these scalar models, 
going beyond the ultra-local approximation without using a 
lattice regularization procedure. Using the strong-coupling perturbative 
expansion we show how it is possible to compute the renormalized vacuum 
energy of a self-interacting scalar field, going beyond the one-loop level.

\footnotetext[1]{e-mail:\,\,nfuxsvai@cbpf.br}

PACS numbers:03.70+k,04.62.+v

\end{titlepage}
\newpage\baselineskip .37in
\section{Introduction}
$\,\,\,\,\,\,\,$ The purpose of this paper is twofold. 
The first is to study the structure of the singularities of the 
perturbative series in different perturbatively renormalizable 
models in field theory. Our method is based in the fact that these 
non-perturbative results can not be obtained using the weak coupling perturbative expansion, and a different  
perturbative expansion is mandatory. Thus, 
in the context of the strong-coupling perturbative expansion,
we investigate the analytic structure of the ultra-local 
generating functionals for the $(\lambda\varphi^{4})_{d}$ model 
and quantum electrodynamics 
in the complex coupling constant plane. The second is  
to discuss how the strong-coupling perturbative expansion 
can be used as an alternative method to compute the renormalized 
free energy or the vacuum energy associated 
with self-interacting fields, going beyond the one-loop level. 
To implement such method we have to show that it is possible to obtain 
renormalizable Schwinger 
functionals associated with scalar 
models going beyond the ultra-local approximation. 
This is performed using an analytic regularization 
procedure. This work is a natural extension of the program 
developed by 
Klauder \cite{jrk}, Rivers \cite{rivers} and others who have been 
studying the strong-coupling expansion and the 
ultra-local generating functional in different   
scalar infrared-free models in field theory.

The perturbative renormalization approach 
in quantum field theory is an 
algorithm where, starting from the Feynman diagrammatic representation 
of the perturbative series, two different steps are usually performed.
In the first step, we control all the ultraviolet divergences of the theory, 
using a procedure to obtain well-defined expressions for each Feynman diagram.
In the second step we have to implement a renormalization prescription where the divergent part of each Feynman diagram is canceled out by a suitable counterterm. For complete reviews
of this program see for example 
Ref. \cite{t Hooft} or Ref. \cite{colins}. 
Concerning the first step, there are different ways to transform 
the Feynman diagrams in well-defined finite 
quantities. The simplest way is to modify the field theory at short 
distances by introducing a sharp cut-off in momentum 
space, or a more elaborate version, as the  
Pauli-Villars regularization \cite{pauli}. 
In this second case we simply modify the propagator for large momentum. 
We can also employ a lattice regularization method in which we 
replace the continuum Euclidean space by a hypercubic 
lattice, with lattice spacing $a$. 
It is clear that the introduction of a lattice provides 
a cut-off in momentum space  
of the order of the inverse of the lattice spacing $a$.
Finally, a more convenient regularization procedure is  
dimensional regularization \cite{dim1} \cite{dim2} \cite{dim3} \cite{dim4}, 
which is particulary well suited to deal with abelian and 
non-abelian gauge theories.

In the second step, to implement the renormalization procedure, 
using dimensional regularization for example, we use the fact that 
the ultraviolet divergences of Feynman diagrams appear as 
poles of some function defined in a complex plane. 
Then the perturbative renormalization 
is performed by the cancelation of the principal 
part of the Laurent series of the analytic regularized expressions.
This cancelation is done introducing counterterms in the theory.
There are different ways to disregard the 
divergent part of each Feynman diagram. For example, we can use the minimal 
subtraction scheme (MS) where the counterterms just cancel the 
principal part of the 
Laurent series of the analytic regularized expressions, or 
any different renormalization 
scheme. The arbitrariness of the method employed must be cured 
by the renormalization group equations \cite{gell} \cite{call} \cite{sym}.

In this framework, field theory models are classified as 
perturbatively super-renormalizable, renormalizable and 
non-renormalizable. 
In the standard weak-coupling perturbative expansion,
the fundamental difference between a perturbatively renormalizable and 
a non-renormalizable model in field theory is given by the 
following property. 
In a non-renormalizable model,  
the usual renormalization procedure used to 
remove the infinities that arise in the usual perturbative 
expansion introduces infinitely many new empirical parameters in the 
theory. This comes from the fact that 
we need to specify the finite part of infinitely many counterterms. 
Consequently, the predictibility or the physical consistency 
of non-renormalizable models is missing.
Thus, renormalizability of field theory models  
provides a valuable constraint on new theories. See, for instance, the discussion in Ref. \cite{cooper}.
Nevertheless, there are some non-renormalizable 
models where it is possible to construct 
a physically sensible version of the theory. A well-known example 
is the Gross-Neveu model, which 
is not renormalizable in the usual sense. However, this model is 
renormalizable in the $\frac{1}{N}$ expansion, for $d<4$ 
\cite{neveu} \cite{parisi1}. We would like to mention that there is a 
alternative approach to deal with non-renormalizable model in 
field theory, using the idea of effective 
field theories \cite{efe} \cite{mario1}, but in this paper we will 
not discuss these issues.

In the case of perturbatively renormalizable models, although the 
renormalization procedure can be implemented in a mathematicaly 
consistent way, it is still not clear how the renormalized 
perturbative series can be summed up in different models 
\cite{dyson} \cite{parisi2} \cite{zinn1}. 
In the literature, there are many results showing that 
the series that we obtain in different perturbative renormalizable 
theories in $d=4$
do not converge for any value of the coupling constants 
of the interacting theories.
Well-known theories with such problems are scalar models with a 
$(\lambda\varphi^{4})$ self-interaction and also 
quantum electrodynamics.
If one tries to perform a partial 
resummation of the perturbative series in both theories, 
Landau poles appear \cite{po} \cite{landau} \cite{po2}.  
In a four-dimensional spacetime, to circumvent the problem of 
non-hermitian Hamiltonians in the infinite cut-off limit, 
the $(\lambda\varphi^{4})_{4}$ model, the $O(N)_{4}$ model, and also 
quantum electrodynamics must be made trivial.

A new step in the development of quantum field 
theory was given by the construction of 
non-abelian gauge field theories and the 
discovery of asymptotic freedom \cite{pol} \cite{gross1} \cite{gross2}
after the construction of the renormalization group equations. 
From the renormalization group equations an important classification of 
different field theory models arises. The models are either 
asymptotically free or IR (infrared) stable.
In the renormalization group approach, the triviality of  
$(\lambda\varphi^{4})_{4}$ model, the $O(N)_{4}$ model and also 
quantum electrodynamics in a four-dimensional spacetime is a 
reflection of the absence of a non-trivial ultraviolet stable 
fixed point in the Callan-Symanzik $\beta$-function.
In the infrared-free theories, for $d=4$,
the problem of the singularities of the connected
three-point and four-point Green's functions for 
quantum electrodynamics and 
$(\lambda\varphi^{4})_{4}$, respectively, is related to the 
following fact: in the framework of the 
weak-coupling expansion, the high frequency 
fluctuations are more strongly coupled than the lower frequency ones.
It is well known that we found a 
completely distinct behavior in a theory where the 
high frequency fluctuations are more weakly coupled than 
the lower frequencies ones. In this situation, at least the 
zero charge problem does not appear.

The unification of statistical mechanics with some models in 
quantum field theory was achieved in progressive steps. 
First Schwinger introduced the idea of Euclidean fields, where the 
classical action must be continued to Euclidean time \cite{ans}.  
Then, Symanzik constructed the Euclidean functional integral 
where the vacuum persistence functional, defined in Minkowski spacetime,
becomes a statistical mechanics average of classical fields 
weighted by a Boltzmann probability \cite{eft} \cite{ef}. 
At the  same time, a deeper insight into our understanding of the 
renormalization procedure 
in different models in field theory was given by the study of the critical 
phenomena and the Wilson version of the 
renormalization group equations \cite{wilson}. 
Further, Osterwalder and Schrader proved that 
for scalar theories, the Euclidean Green's 
functions or the Schwinger functions, which are the 
moments of the Boltzmann measure, 
are equivalent to the Minkowski Green's functions \cite{os}. 
A new step in our understanding 
of the limitation of the perturbative approach in 
quantum field theory was achieved by Aizenman \cite{aiz} and 
Frohlich \cite{fro}.
These authors proved that the $(\lambda\varphi^{4})_{d}$ 
model, with the use of a lattice regularization with nearest 
neighborhood realization of the 
Laplacian, leads to a trivial theory in the continuum limit for 
$d\geq 5$. For $d=4$, with additional assumptions, it is also 
possible to obtain the triviality of the model. For an interesting 
review of the triviality problem in quantum field theory, 
see Ref. \cite{calawaw}.

It is important to point out that some authors claim that 
the triviality of  $(\lambda\varphi^{4})_{d}$ for $d\geq 4$
is an odd result, since for $d=4$
the renormalized perturbative series is non-trivial and 
for $d\geq 5$ the theory is 
perturbatively non-renormalizable. For example, Klauder 
has argued that the triviality of 
$(\lambda\varphi^{4})_{d}$ for $d\geq 4$ is still an 
open problem \cite{kk}. 
Making use of the correspondence principle, this author 
has emphasized that 
the quantization of a non-trivial classical theory 
can not be a non-interacting quantum theory. 
Furthermore, he claims that  
an alternative regularization procedure can give 
a non-trivial theory in the infinite cut-off limit.

Some results going in this direction have been obtained by 
Gallavotti and Rivasseau \cite{ga}.
These authors discuss the scalar $(\lambda\varphi^{4})_{d}$ theory
with more general regularized theories 
where the realization of the Laplacian is not restricted to the nearest neighbours and 
also the presence of antiferromagnetic couplings. They suggested that the 
ultraviolet limit of such lattice regularized field theories is not a 
Gaussian field theory model, which would open  
the possibility to construct 
scalar models with a non-trivial ultraviolet limit. 
We shall emphasize that the ultraviolet behavior 
of the $(\lambda\varphi^{4})_{d}$ model for $d\geq 4$ is a 
strong-coupling problem, 
since, in the weak coupling perturbative expansion, 
the high-order terms of the perturbative series are dominant 
in the large 
cut-off limit, as has been discussed by many authors. 
See, for instance, the 
discussion in Ref. \cite{sokal}. 
Another situation in field theory where the 
weak-coupling perturbative expansion is not appropriate is in the 
large distance behavior of quantum chromodynamics.

In the case of a strong-coupling regime of a theory, we can 
try to obtain rigorous results by the use of constructive field theory,
perform a partial resummation of the Feynman diagrams improving 
the Feynman-Dyson perturbative series, or
perform a different perturbative 
expansion by using the following approaches. 
The first is to introduce auxiliary fields in order to 
disconect the interaction part from the free part of the 
Lagrangian density and then to perform a perturbative expansion 
of the Schwinger functional in inverse powers of the coupling constant
\cite{ward}. For example, the $\frac{1}{N}$ 
expansion, where $N$ is the number of the components of the field in some 
isotopic space (the dimension of the order parameter), is 
a realization of this approach \cite{cjp}. Of course we are still using the 
standard perturbative scheme, performing a perturbative 
expansion with respect to the anharmonic terms of the theory. 
The basic idea of the second approach is to construct a 
formal representation for the generating functional of 
complete Schwinger functions of the theory, treating 
the off-diagonal terms of the Gaussian factor as 
a perturbation about the remaining terms in the functional integral. 
This approach has been called the strong-coupling expansion 
\cite{kovesi} \cite{be1} \cite{parga} \cite{be2} \cite{co} \cite{be3}.

The purpose of this paper is to discuss some of the problems and virtues of  
the strong-coupling expansion in 
different scalar models and also quantum electrodynamics. We first study the 
singularities of the generating functional of complete Schwinger functions 
for the $(\lambda\varphi^{4})_{d}$ model
in the complex coupling constant plane 
by examining the analytic structure of the zero-dimensional generating function. Then, we repeat the analysis for the case of
quantum electrodynamics.
Second we  present two idealized interacting field theory models where 
the weak-coupling expansion cannot be used.
In these cases the strong-coupling expansion
is more adequate to investigate the properties of the models.
It is important to stress that the analytic structure of the 
Schwinger functions (in the weak-coupling perturbative expansion framework) 
and also of the Schwinger functional can be easily obtained using the 
ultra-local approximation derived in the strong-coupling perturbative expansion. 
As we will see, these non-perturbative results do not change if 
we go beyond the ultra-local approximation. We should say at this point that Bender et al \cite{conf} also consider the 
zero-dimensional field theory to obtain 
non-perturbative results in field theory.   
Third, using an analytic regularization procedure,
we discuss briefly how it is possible to obtain a renormalized 
Schwinger functional going beyond the ultra-local approximation.
Finally, we sketch how the strong-coupling perturbative expansion 
can be used as an alternative method to compute the renormalized 
free energy or the vacuum energy associated 
with self-interacting fields, going beyond the one-loop level.

There are three points that we would like to briefly discuss. First, it is
interesting to known whether an ultra-local approximation has 
been used in the literature in other contexts. In the Laudau theory of continuous phase transition, which 
reproduces the mean-field exponents, we have a simplified version of the 
ultra-local approximation, since in the partition function we 
drop the gradient
term and the sum will be dominated by its largest term. 
Further, Landau and Ginzburg modified the the original Landau theory by 
introducing the gradient term into the energy density 
that discourages rapid fluctuations in the order parameter \cite{phase}. 
Before continuing we would like to point out that the strong-coupling perturbative expansion is quite similar to the high-temperature 
series expansion in statistical mechanics. A similar idea is the Mayer expansion, a method for carrying out the cluster expansion, introduced into 
quantum field theory by Symanzik \cite{mayer}.  Finally, in the 
study of critical phenomena using lattice simulation, there is an 
analog of the strong-coupling perturbative expansion: the hopping parameter
expansion, where the perturbative expansion starts from the 
disordered lattice system \cite{reisz}. For a complete review of the 
linked cluster expansion, see for example Ref. \cite{lce}.

Second, it is interesting to point out that the study of the ultra-local 
field theory models can also bring some insights into 
non-renormalizable field theory models. As a simple example, let 
us suppose a massive abelian vector field $W_{\mu}(x)$ coupled with fermions. 
Using a Fourier representation for the Euclidean two-point Schwinger function 
associated with the massive vector field $W_{\mu}(x)$, the Fourier 
transforms $G_{\mu\nu}(k)$ are given by
$$
G_{\mu\nu}(k)=\left(\delta_{\mu\nu}+\frac{k_{\mu}\,
k_{\nu}}{m^{2}}\right)\frac{1}{k^{2}+m^{2}}.
$$
The Fourier transform  $G_{\mu\nu}(k)$
of the two-point Schwinger function does not vanish in 
the ultraviolet limit $k\rightarrow \infty$ and its large Euclidean 
momentum behavior is roughly $m^{-2}$. Consequently the behavior 
in this limit is similar to the behavior 
for the two-point Schwinger function in the ultra-local  $(\lambda\varphi^{4})_{d}$ model.

Finally, we would like to call 
the attention of the reader that the study of the analytic structure 
of theories in the complex coupling constant 
plane has been used by many authors in quantum field theory. 
It is well known that the behavior of the standard  
perturbative series in powers of the coupling constant at large 
order is related to the analytic structure 
of the partition function in a neighborhhood of the origin in 
the complex coupling constant plane. For example, 
Bender and Wu \cite{wu} 
studied the anharmonic oscillator and pointed out that there is a relation 
between the $n^{th}$ Rayleigh-Schrodinger coefficients and the 
lifetime of the unstable states of a negative coupling constant anharmonic oscillator.

The organization of the paper is as follows: In section II 
we discuss the standard weak-coupling expansion for 
the $(\lambda\varphi^{4})_{d}$ model. In section III we 
discuss the strong-coupling expansion for 
the $(\lambda\varphi^{4})_{d}$ model.
In section IV we perform the study of the 
analytic structure of the  
zero-dimensional $(\lambda\varphi^{4})$ model in the 
complex coupling constant plane. 
In section V we perform the study of the 
analytic structure of the  
zero-dimensional quantum electrodynamics in the 
complex coupling constant plane.
In section VI we present idealized models 
where the strong-coupling 
perturbative expansion must be used.
In section VII we sketch how it is possible to go beyond 
the ultra-local approximation in general scalar models.
In section VIII we show how it is possible to compute the 
renormalized vacuum energy associated with a self-interacting 
scalar field in the presence of macroscopic structures 
going beyond the one-loop level. Finally, section IX 
contains our conclusions.  
To simplify the calculations we assume the units to be such that 
$\hbar=c=1$, and also all the physical quantities are dimensionless.
Consequently it is convenient to introduce 
an arbitrary parameter $\mu$ with dimension of
mass to define all the dimensionless physical quantities.    
Thus, in the paper we are using dimensionless cartesian coordinates $(x^{\alpha}=\mu\,x'^{\alpha})$, where $x'^{\alpha}$ are the usual 
coordinates with dimension of length.

\section{Weak coupling perturbative expansion for 
the scalar $(\lambda\varphi^{4})_{d}$ model}\

Let us consider a neutral scalar field with a 
$(\lambda\varphi^{4})$ self-interaction, defined in a d-dimensional 
Minkowski spacetime.
The vacuum persistence functional is   
the generating functional of all 
vacuum expectation value of time-ordered products of the theory.
The Euclidean field theory can be obtained by analytic 
continuation to imaginary time supported by the positive 
energy condition for the relativistic field theory.
In the Euclidean field theory, we have the Euclidean counterpart 
for the vacuum persistence functional, that is, the generating 
functional of complete Schwinger functions.
The $(\lambda\varphi^{4})_{d}$ Euclidean theory 
is defined by these Euclidean Green's functions. 
The Euclidean generating functional $Z(h)$ is formally 
defined by the following functional integral:  
\begin{equation}
Z(h)=\int [d\varphi]\,\, \exp\left(-S_{0}-S_{I}+ \int d^{d}x\, 
h(x)\varphi(x)\right),
\label{1}
\end{equation}
where the action that usualy describes a free scalar field is 
\begin{equation}
S_{0}(\varphi)=\int d^{d}x\, \left(\frac{1}{2}
(\partial\varphi)^{2}+\frac{1}{2}
m_{0}^{2}\,\varphi^{2}(x)\right),
\label{2}
\end{equation}
and the interacting part, defined by the non-Gaussian contribution, 
is 
\begin{equation}
S_{I}(\varphi)= \int d^{d}x\,\frac{g_{0}}{4!}
\,\varphi^{4}(x).
\label{3}
\end{equation} 
In Eq.(\ref{1}), $[d\varphi]$ is 
a translational invariant measure, formally given by 
$[d\varphi]=\prod_{x} d\varphi(x)$.
The terms $g_{0}$ and $m_{0}^{2}$ are respectivelly the 
bare coupling constant and mass squared of the model.
Finally, $h(x)$ is a smooth function that we introduce 
to generate the Schwinger functions of the 
theory by functional derivatives. Note that we are using the same notation
for functionals and functions, for example $Z(h)$ instead the usual notation
$Z[h]$.

In the weak-coupling perturbative expansion, which is the conventional procedure, we 
perform a formal perturbative expansion with respect to the non-Gaussian
terms of the action. As a consequence of this formal expansion,
all the $n$-point unrenormalized Schwinger functions are 
expressed in a powers series of the bare coupling 
constant $g_{0}$. Let us summarize how to perform the weak-coupling 
perturbative expansion in the 
$(\lambda\varphi^{4})_{d}$ theory, and briefly discuss also 
the divergent behavior of the perturbative series. 
The Gaussian functional integral $Z_{0}(h)$ associated with the 
Euclidean generating functional $Z(h)$ is  
\begin{equation}
Z_{0}(h)=\,{\cal{N}}\int\,[d\varphi]\,\,
\exp\left(-\frac{1}{2}\varphi\,K\,\varphi+h\varphi\right).
\label{zz1}
\end{equation}
We are using the compact notation of 
Zinn-Justin \cite{livron} and each term in Eq.(\ref{zz1}) is 
given by
\begin{equation}
\varphi\,K\,\varphi=\int d^{d}x\,\int d^{d}y\, \varphi(x)K(m_{0};\,x,y)\varphi(y),
\label{zz2}
\end{equation}
and
\begin{equation}
h\varphi=\int d^{d}x\, \varphi(x)h(x).
\label{zz3}
\end{equation}
As usual ${\cal{N}}$ is a normalization factor and 
the symmetric kernel $K(m_{0};x,y)$ is defined by 
\begin{equation}
K(m_{0};x,y)=(-\Delta+m^{2}_{0}\,)\,\delta^{d}(x-y),
\label{benar}
\end{equation}
where $\Delta$ denotes the Laplacian in $R^{d}$.
As usual, the normalization factor is defined using the condition 
$Z_{0}(h)|_{h=0}=1$. Therefore  
${\cal{N}}=\left[\mbox{det}(-\Delta+m_{0}^{2})\right]^{\frac{1}{2}}$
but, in the following, we are absorbing this normalization factor in the
functional measure. 
It is convenient to introduce the inverse kernel, i.e.
the free two-point Schwinger function 
$G_{0}(m_{0};x-y)$ which satisfies  the 
identity  
\begin{equation}
\int d^{d}z\,G_{0}(m_{0};x-z)K(m_{0};z-y)=\delta^{d}(x-y).   
\label{zz4}
\end{equation}
Since 
Eq.(\ref{zz1}) is a Gaussian functional integral, 
simple manipulations, performing only Gaussian integrals, give 
\begin{equation}
\int [d\varphi]\,\, \exp\left(-S_{0}+\int d^{d}x\, h(x)\varphi(x)\right)=
\exp\left(\frac{1}{2}\int d^{d}x\, \int d^{d}y\,\,h(x)\,G_{0}(m_{0};x-y)h(y)\right).
\label{6}
\end{equation}
Therefore, we have an expression for 
$Z_{0}(h)$ in terms of the inverse kernel 
$G_{0}(m_{0};x-y)$, i.e., in terms of the free two-point Schwinger function.

This construction is fundamental to perform the Feynman-Dyson
weak-coupling perturbative expansion
with the Feynman diagramatic representation of the perturbative series.
Using Eqs.(\ref{1}), (\ref{2}), and (\ref{6}),
we are able to write the generating functional of all 
bare Schwinger functions $Z(h)$ as 
\begin{equation}
Z(h)= \exp\left(-\int d^{d}x\,{\cal{L}_I}(\frac{\delta}{\delta h})\right)
\exp\left(\frac{1}{2}
\int d^{d}x\,\int d^{d}y\,\, h(x)\,G_{0}(m_{0};x-y)h(y)\right),
\label{8}
\end{equation}
where ${\cal{L}_I}$ is defined by the non-Gaussian contribution 
to the action.
Consequently, in the conventional perturbative expansion, 
the generating functional of complete Schwinger functions 
is formally given by 
the following infinite series
\begin{equation}
Z(h)=\sum_{n=0}^{\infty}\frac{(-1)^{n}}{n!}
\left(\int d^{d}x \,\,{\cal{L}_{I}}(\frac{\delta}{\delta h})\right)^{n}
\exp\left(\frac{1}{2}\int d^{d}x\, \int d^{d}y\,\,h(x)\,G_{0}(m_{0};x-y)h(y)\right).
\label{uta}
\end{equation}
To generate all the $n$-point Schwinger functions we have only
to perform a suitable number of functional differentiations in $Z(h)$ with 
respect to the source $h(x)$ and set the source to zero in the end. 
The bare $n$-point Schwinger functions are defined by
\begin{equation}
G_{n}(x_{1},x_{2},..,x_{n})=Z^{-1}(h=0)\left[\frac{\delta}
{\delta h(x_{1})}...\frac{\delta}{\delta h(x_{n})}\,Z(h)\right]|_{h=0}.
\label{z1}
\end{equation}
This general method can be used 
to derive the weak-coupling perturbative expansion in different theories. 
Observe that it is possible to generalize this formalism including the 
product of composite sources \cite{cjt1} \cite{cjt2} \cite{cjt3}, but in this 
paper we limit ourselves to models without composite operators.
To generate only the connected diagrams $G_{n}^{(c)}(x_{1},..,x_{n})$, let us
consider the generating functional of the connected 
Schwinger functions (the free energy functional), 
defined as $F(h)=\ln\, Z(h)$. 
The functional Taylor expansion of 
the generating functional of connected Schwinger functions is 
\begin{equation}
F(h)=\sum_{n=1}^{\infty}\frac{1}{n!}\int \prod_{k=1}^{n}d^{d}x_{k}\prod_{k=1}^{n}
h(x_{k})\,G_{n}^{(c)}(x_{1},..,x_{n}).
\label{9}
\end{equation}  
The functional Taylor expansion of the generating functional of all 
Schwinger functions $Z(h)$ in powers of the coupling constant $g_{0}$ is
\begin{equation}
Z(h)={\cal{N}}\left(1+\sum_{k=1}^{\infty}g_{0}^{k}\,A_{k}(h)\right),
\label{10}
\end{equation}
where $A_{k}(h)$ are perturbative coefficients.
After a regularization and renormalization procedure it is possible 
to show that any physically measurable quantity $f(g)$ can be expanded in 
power series defined by
\begin{equation}
f(g)=\sum_{k=0}^{\infty}\,f_{k}\,g^{k},
\label{11}
\end{equation}
where $g$ is the renormalized coupling constant and $f_{k}$ are perturbative 
coefficients.

It is well known that in general the series that we obtain from 
perturbatively renormalizable theories are divergent. 
Consequently the perturbative renormalization method is 
not enough to obtain well defined physical quantities.
For example, for $P(\varphi)_{2}$ 
the renormalized perturbative series for any connected Schwinger function 
that can be obtained by a Wick ordering is divergent \cite{p2}.
For the $(\lambda\varphi^{4})_{3}$ model a similar divergent 
behavior was proved by de Calan and Rivasseau \cite{p3}.
Although in general the series that we obtain from perturbatively 
renormalizable theories are divergent, there is an impressive agreement  
of the theoretical results with the experiments, 
when someone use the first terms of the weak-coupling perturbative expansion
to extract predictable results. 
Therefore, these divergent series shall be an asymptotic expansion of 
the solutions of the theories. 
In other words, in a specific theory  
even though the renormalized perturbative series diverges, a finite number of terms of 
the series is still a good approximation of 
the functions in question.
Now we need a tool for obtaining the functions of the theory 
from the divergent series. 
The Borel resummation is this tool that allows us to obtain 
the solutions of the theory from these divergent series \cite{knop}.   
Let us discuss with more details the asymptotic expansion of a function and  
Borel summability. Consider a function $f(z)$ defined in the complex plane for large $z$. 
The formal series $\sum_{n=0}^{\infty}\,a_{n}z^{-n}$, 
which need not converge for 
any value of $z$, is called the 
asymptotic expansion or asymptotic representation of the 
function $f(z)$ if defining
\begin{equation}
i) \,\,S_{N}(z)=\sum_{n=1}^{N}\, a_{n}z^{-n}
\label{ass1}
\end{equation}
\begin{equation}
ii) \,\,R_{N}(z)=z^{N}|f(z)-S_{N}(z)|,
\label{ass2}
\end{equation}
we have $\lim_{|z|\rightarrow\infty}R_{N}(z)=0$, for every fixed $N$.
There is a similar definition of the asymptotic expansion of a function near zero, involving 
series of the kind $\sum_{n=0}^{\infty}\,a_{n}z^{n}$. 
Again, this series is the asymptotic representation of $f(z)$, i.e.,
$f(z)\sim \sum_{n=0}^{\infty}\,a_{n}z^{n}$ if for a small $z$ we still 
have $(i)$ and in $(ii)$ we have 
$R_{N}(z)=z^{-N}|f(z)-S_{N}(z)|$ and for every fixed $N$ we have 
$lim_{|z|\rightarrow\,0}R_{N}(z)=0$.
From the above definition there 
are two main questions. The first is the question whether a 
function under consideration possesses 
an asymptotic expansion, which we call the expansion problem. 
There is also the question of how the 
function is to be found, which is represented by a 
given asymptotic expansion, that we 
call the summation problem. 
Note that any function can have only one asymptotic expansion, 
or we can show that 
the function in question has no asymptotic expansion. 
Suppose $f(x)=\exp(-x),\,\,x>0$. It is clear that 
all the coefficients of the asymptotic expansion are 
zero since $x^{k}\exp(-x)\rightarrow 0$ for every 
$k\geq 0$ when $x\rightarrow\infty$.  This simple result shows 
that different functions may have 
the same asymptotic expansion. For example if $h(z)$ has 
an asymptotic representation it is clear 
that $h(z)+\exp(-z)$ or $h(z)+a\exp(-bz)$ for $b>0$ have 
the same asymptotic representation. 
Now, consider a function $f(z)$ which has the asymptotic expansion 
in a region of the complex plane,
defined by a divergent series. Thus we have 
\begin{equation}
f(z)\sim\,\sum_{k=0}^{\infty}f_{k}\,z^{k}.
\label{b1}
\end{equation}
The Borel transform of $f(z)$, called 
$B_{f}(z)$, is defined as
\begin{equation}
B_{f}(z)=\sum_{k=0}^{\infty}\frac{1}{k!}f_{k}\,z^{k}.
\label{b2}
\end{equation}
The key point is that the Borel transform may converge even if the series 
is divergent. The Borel resummation of the series is obtained 
applying  the inverse Borel transform on $B_{f}(z)$, given by
\begin{equation}
f(z)=\int_{0}^{\infty}\,\exp(-t)\,B_{f}(zt)\,dt.
\label{b3}
\end{equation}
This construct is an 
indispensable tool to recover a function from its asymptotic expansion in quantum 
field theory. A pedagogical discussion of the application of the 
Borel transform in perturbation theory can be found in Ref. \cite{parisi}.
It is not difficult to repeat this construction for the n-point Schwinger 
function of any renormalizable theory. 
If the perturbative series of the 2n-point renormalized
Schwinger functions does not 
converge, as these series must be an asymptotic expansion 
for the solutions of our theory, the Borel 
resummation method can be used to recover the solutions.
In the standard perturbative expansion we express the 
2n-point renormalized
Schwinger functions as the following power series in the 
renormalized coupling constant $g$: 
\begin{equation}
G_{2n}(g;\,x_{1},x_{2},..,x_{2n})\sim\,
\sum_{k=0}^{\infty}\,g^{k}\,G_{2n}^{(k)}(x_{1},x_{2},..,x_{2n}).
\label{b4}
\end{equation}
Let us define the Borel transform of the n-point Schwinger function by
\begin{equation}
G_{2n}(\tau;\,x_{1},x_{2},..,x_{2n})=\sum_{k=0}^{\infty}\,
\frac{\tau^{k}}{k!}\,G_{2n}^{(k)}(x_{1},x_{2},..,x_{2n}),
\label{b5}
\end{equation}
and it is clear that from the inverse Borel transform we have 
\begin{equation}
G_{2n}(g;\,x_{1},x_{2},..,x_{2n})=\frac{1}{g}\int_{0}
^{\infty}d\tau\,\exp(-\frac{\tau}{g})\,
G_{2n}(\tau;\,x_{1},x_{2},..,x_{2n}).
\label{b6}
\end{equation}
The series which defines $G_{2n}(\tau;\,x_{1},x_{2},..,x_{2n})$ in Eq. (\ref{b5}) has much 
better convergence properties than
the original series of $G_{2n}(g;\,x_{1},x_{2},..,x_{2n})$ in Eq.(\ref{b4}). 
Eckmann et al.~\cite{eck}
obtained the Schwinger functions of the  
$(\lambda\varphi^{4})_{2}$ model from the divergent perturbative series 
using Borel resummation method. The extension of this result to 
$(\lambda\varphi^{4})_{3}$ model was obtained by Magnen and Seneor \cite{eck1}.

Note that there are some situations where the Borel resummation method can not
be implemented. This happens, for example, when the Borel transform has 
singularities in the real line. These singularities are  
related to non-perturbative effects which are not apparent 
in the weak-coupling perturbative expansion \cite{lipatov}. 
In the absence of singularities in the positive axis 
of the Borel transform,  
the Borel resummation method is a powerful way to extract results 
from a divergent series. See also Ref. \cite{kuri1} and Ref. \cite{to}.

\section{The strong-coupling perturbative expansion for 
the scalar $(\lambda\varphi^{4})_{d}$ model}\

Many phenomena in quantum field theory cannot be described in 
the framework of the weak-coupling perturbative expansion. For example,
in the case of the strong-coupling regime of a theory, the perturbative 
expansion in powers of the coupling constant is unrealiable.
From now on, we shall study the strong-coupling perturbative expansion
in different models in field theory. As we discussed
this perturbative expansion may be used 
in the strong-coupling regime of a model, as for example   
the ultraviolet limit of a non-asymptoticaly free model or as an
alternative expansion in models with non-polynomial Lagrangian interaction. 
In the first situation, or in the case of non-renormalizable theories,
it is imperative to investigate an alternative perturbation expansion. Before continue, we would like to call the attention of the reader 
that a alternative perturbative program for dealing with 
non-renormalizable theories, has been 
developed by Klauder and others \cite{novoc2} \cite{novoc1} 
\cite{novoc3} \cite{novoc4}. Klauder proposed a non-canonical formulation 
for the quantization  of $(\lambda\varphi^{p})_{4}$ model $(p>4)$ using a non-translational invariant functional measure.
To support this approach he observes that there are many 
situations where an infinitesimal perturbation causes a discontinuous change 
in the eigenfunctions and eigenvalues associated with a Hamiltonian system
\cite{ezawa} \cite{kay}.  For example in the 
$(\lambda\varphi^{p})_{4}$ model, if $p\leq 4$ 
the theory is renormalizable and field configurations which have a finite free 
action also give a finite contribution for the interaction term. For 
$p>4$ this does not happen: the free field situation can not be obtained when 
$\lambda\rightarrow 0^{+}$ and this limit is the pseudo-free solution. 
In this paper we decided to follow a more conventional treatment. 
We use an alternative perturbative expansion, but assuming 
that the measure in the functional integral is translational 
invariant instead of using a scale covariant measure.

The lesson that we have from these discussions is that 
if someone decides to perform a perturbative expansion of a 
strongly-coupled theory, then a  
resummation of the weak-coupling perturbative series to obtain 
non-perturbative results is necessary.
An alternative procedure is not to use the conventional 
perturbation theory around the Gaussian-free theory.  
Consequently, we now turn to the alternative expansion 
that has been called in the literature the strong-coupling perturbative expansion. The basic idea of this approach 
is to treat off-diagonal terms of the 
Gaussian factor as a perturbation about the remaining diagonal 
terms in the integral. 
Although we are studying only scalar models, the extension to 
higher spin fields is straightforward. See for example the 
discussion given in Ref. \cite{efimov}.

Let us suppose a compact Euclidean space with or withouth a boundary.
An equivalent possibility is to work in an unbounded Euclidean space but 
assume that the functional integral is taken over field configurations that vanish at large Euclidean distances. Let us 
suppose that there exists an eliptic, semi-positive, and self-adjoint 
differential 
operator $D$ acting on scalar functions on the  Euclidean space.
The usual examples are
$D=(-\Delta)$ and $D=(-\Delta+m_{0}^{2})$. The kernel $K(m_{0};\, x-y)$ 
is given by $K(m_{0};\, x-y)=D\,\delta^{d}(x-y)$. 
Let us study first the self-interacting 
$(\lambda\varphi^{4})_{d}$ model. Thus we have:
\begin{equation}
{\cal{L}_{I}}(g_{0}; \varphi)=\frac{g_{0}}{4!}\varphi^{4}(x).
\end{equation}
Treating the off-diagonal terms of the 
Gaussian factor as a perturbation about the remaining terms in the integral,
we get a formal expression for the generating functional of complete 
Schwinger functions $Z(h)$:
\begin{equation}
Z(h)=\exp\left(-\frac{1}{2}\int d^{d}x\,\int d^{d}y\,
\frac{\delta}{\delta h(x)}K(m_{0};x-y)
\frac{\delta}{\delta h(y)}\right)\,Q_{0}(h),
\label{12}
\end{equation}
where the ultra-local generating functional $Q_{0}(h)$ is defined by 
\begin{equation}
Q_{0}(h)={\cal{N}}\int [d\varphi]\,
\exp\left(\int d^{d}x\,(\,-\frac{g_{0}}{4!}\varphi^{4}(x)+
h(x)\varphi(x))\right).
\label{13}
\end{equation}
The factor ${\cal{N}}$ is a normalization that can be found using that 
$Q_{0}(h)|_{h=0}=1$. 
Observe that all the non-derivative terms in the original action appear
in the functional integral that defines $Q_{0}(h)$. 
As we discussed in section II, in the  $(\lambda\varphi^{4})_{d}$ model, 
the kernel is defined by Eq.(\ref{benar}): 
$K(m_{0};x-y)=(-\Delta+m^{2}_{0})\,\delta^{d}(x-y)$. 
At this point it is convenient to 
consider $h(x)$ to be complex. Consequently
$h(x)=\mbox{Re}(h)+i\,\mbox{Im}(h)$ (we are concerned with the case  
$\mbox{Re}(h)=0$). It should be noted that 
although the functional integral $Q_{0}(h)$ is not 
a product of Gaussian integrals, it can be viewed formally as an 
infinite product of ordinary integrals, one for each point of the 
d-dimensional Euclidean space.
The fundamental problem of the strong-coupling 
expansion is how to construct non-Gaussian 
measures to define the 
Schwinger functional. It is important to point out that 
the solution of this problem would  
allow us to deal with non-renormalizable models
in the weak-coupling expansion framework.

The expansion of the exponential term on Eq.(\ref{12})
gives a formal expansion of $Z(h)$
as a perturbative series 
in the following form: 
\begin{equation}
Z(h)=\sum_{i=0}^{\infty}\,Z^{(i)}(h),
\label{q1aa}
\end{equation}
where the two first terms of the perturbative series are respectively the 
ultra-local generating functional $Q_{0}(h)$ and $Z^{(1)}(h)$ defined by
%
%
\begin{equation}
Z^{(1)}(h)=\left(-\frac{1}{2}\int 
d^{d}x\,\int d^{d}y\,\frac{\delta}{\delta h(x)}K(m_{0};x-y)
\frac{\delta}{\delta h(y)}\right)\,Q_{0}(h).
\label{q3}
\end{equation}
The main difference from the standard perturbative expansion is that 
we have an expansion of the generating functional of complete Schwinger functions in inverse powers of the coupling constant. We are developing our 
perturbative expansion around the static ultra-local functional $Q_{0}(h)$ 
\cite{kla} \cite{ca} \cite{kla2} \cite{meni}.  
Fields defined in different points of the Euclidean  
space are decoupled in the ultra-local approximation 
since the gradient term is dropped.

As we stressed, although the strong-coupling expansion 
is very inconvenient for pratical calculations in 
the continuum $R^{d}$ Euclidean space, it is very natural in the lattice. 
The technical 
problems that we have to deal with in the 
continuum Euclidean space are the following: first we 
have to define non-Gaussian functional measures; 
second we have to regularize and renormalize 
the Schwinger functions obtained from the generating functional, 
going beyond the ultra-local approximation. 
We would like to stress that
we will not use a lattice structure 
of the Euclidean space 
as a regulator to implement the  renormalization program. Instead, we 
use the lattice structure only to define what we mean by the 
ultra-local generating functional $Q_{0}(h)$.

In this paper
we are not interested in regularizing the complete 
series of the strong-coupling perturbative expansion.
We are interested first, in the analytic structure of the 
Schwinger functional in the 
complex coupling constant plane. We will show that the 
zero-dimensional generating function of  
the $\lambda\varphi^{4}$ model has 
a branch point singularity. This kind of singularity 
in the complex coupling constant plane 
is also present in the ultra-local 
quantum electrodynamics \cite{mario}.  
Our second interest is to study scalar models where the 
use of the strong-coupling expansion is mandatory. Third,
we briefly discuss how it is possible to use an analytic regularization 
procedure, in order to obtain a renormalized  Schwinger
functional, going beyond the ultra-local approximation.
Finally, we sketch how the strong-coupling perturbative expansion 
can be used as an alternative method to compute the renormalized 
free energy or the vacuum energy associated 
with self-interacting fields, going beyond the one-loop level.

Using the fact that the functional integral which defines 
$Z(h)$ is invariant 
with respect to the choice of the quadratic part,
let us consider a slightly modification of the strong-coupling expansion.  
We split the quadratic part in the functional integral 
proportional to the mass squared in two parts: the off-diagonal 
terms of the Gaussian factor and the ultra-local generating functional. 
The new generating functional of the complete Schwinger functions will be defined by the following functional integral
\begin{equation}
Z(h)=\exp\left(-\frac{1}{2}\int d^{d}x\,\int d^{d}y\frac{\delta}{\delta 
h(x)}K(m_{0},\sigma;x-y)
\frac{\delta}{\delta h(y)}\right)\,Q_{0}(\sigma;\,h),
\label{18}
\end{equation}
where  $Q_{0}(\sigma;h)$,
the new ultra-local functional integral, is given by  
\begin{equation}
Q_{0}(\sigma;h)={\cal{N}}\int [d\varphi]\,\exp\left
(\int d^{d}x\,(-\frac{1}{2}\sigma\, m_{0}^{2}\,\varphi^{2}(x)-
\,\frac{g_{0}}{4!}\varphi^{4}(x)
+h(x)\varphi(x))\right).
\label{19}
\end{equation}
and the new kernel $K(m_{0},\sigma;x-y)$ is defined by 
\begin{equation}
K(m_{0},\sigma;x-y)=\left(-\Delta+(1-\sigma)m_{0}^{2}\,\right)\delta^{d}(x-y), 
\label{benar2}
\end{equation}
where $\sigma$ is a complex parameter defined in the 
region $0\leq\,Re(\sigma)\leq\,1$. 
The choice of a suitable  $\sigma$ will simplify our 
calculations in some situations.

In the next section we study the analytic structure of the 
ultra-local $(\lambda\varphi^{4})_{d}$ theory. 
The divergences that appear in the formal representation 
of the Schwinger functional $Z(h)$ are of two kinds. 
The first kind is related to the infinite volume and continuum 
hypotesis of the Euclidean space and this divergence can be controled by the 
introduction of a box and a regulator function.
The second kind of divergences is related to the functional form of 
the non-Gaussian part of the action and appears as a divergent perturbative series. We are first concerned with this second kind of divergences.

The study of the ultra-local model in different
field theories will clarify the structure of the singularities for the 
perturbative series, and also the structure of the singularities 
for the $n$-point Schwinger functions in the complex coupling constant plane. 
We first investigate the analytic structure of 
the $(\lambda\varphi^{4})_{d}$ ultra-local model, in the complex coupling 
constant plane.

\section{The analytic structure of the ultra-local 
 $(\lambda\varphi^{4})_{d}$ model.}\

The aim of this section is to analyze the analytic structure of the 
zero-dimensional $\lambda\varphi^{4}$ model in the complex coupling 
constant $g_{0}$ plane. As we discussed before, the first term of the 
strong-coupling expansion of $Z(h)$ is exactly the 
ultra-local model \cite{kla} 
\cite{ca} 
\cite{kla2}  
\cite{do} \cite{cant}, also called the static independent value model. 
Since we can interpret 
the ultra-local model as an infinite product of ordinary integrals, let us 
introduce a Euclidean lattice and analyse the generating function defined 
in each point of the Euclidean lattice given by
\begin{equation}
z(m_{0},g_{0};h)=
\frac{1}{\sqrt{2\pi}}
\int_{-\infty}^
{\infty}d\varphi\,\exp\left(-\frac{1}{2}m_{0}^{2}\,\varphi^{2}-
\frac{g_{0}}{4!}\varphi^{4}+h\varphi\right),
\label{source}
\end{equation}
where for simplicity we are assuming $\sigma=1$. 
The generating function in the absence of external 
source is defined by $z(m_{0},g_{0};h)|_{h=0}\equiv z_{0}(m_{0},g_{0})$.
Consequently, our aim is to analyse the following integral with a
quartic probability distribution in which the zero-dimensional 
partition function $z_{0}(m_{0},g_{0})$ is given by
\begin{equation}
z_{0}(m_{0},g_{0})=\frac{1}{\sqrt{2\pi}}\int_{-\infty}^
{\infty}d\varphi\,\exp\left(-\frac{1}{2}m_{0}^{2}\,\varphi^{2}-
\frac{g_{0}}{4!}\varphi^{4}\right).
\label{21}
\end{equation}

Note that this integral is well defined for $\mbox{Re}\,\,g_{0} \geq 0$.
As the exponential power series is convergent everywhere, we 
may write 
\begin{equation}
z_{0}(m_{0},g_{0})=\frac{1}{\sqrt{2\pi}}\int_{-\infty}^
{\infty}d\varphi\,
\sum_{k=0} ^\infty
  \exp\left(-\frac{1}{2}m_{0}^{2}\,\varphi^{2}\right)
\,  (-\frac{g_0}{4!})^k\frac{\varphi^{4k}}{k!}.
\label{21b}
\end{equation}
The above series is not  uniformly convergent. Therefore
we can not interchange the integration and the summation. 
Nevertheless let us perform this interchange, integrate 
term by term, and get a formal series $z^{(1)}(m_{0},g_{0})$.
This formal series is the asymptotic expansion for  $z_{0}(m_{0},g_{0})$.
Thus we have $z_{0}(m_{0},g_{0})\sim z^{(1)}(m_{0},g_{0})$ and 
choosing $m_{0}^{2}=1$ it is not difficult to show that 
\begin{equation}
z^{(1)}(m_{0},g_{0})|_{m_{0}^{2}=1}=\sum_{k=0}^{\infty}\,(-g_{0})^{k}c_{k},
\label{22}
\end{equation}
where the coefficients $c_{k}$ are given by
$c_{k}=\frac{(4k-1)!!}{(4!)^{k}k!}$. 
The formal series $z^{(1)}(m_{0},g_{0})$ 
is divergent for any non-nule value of $g_{0}$. 
The asymptotic expansion for the zero-dimensional
partition function given by $z^{(1)}(m_{0},g_{0})$ 
has the contribution from the vacuum diagrams \cite{id} \cite{bw}, 
and each coefficient $c_{k}$ is given by the sum of symmetry factors over all 
diagrams of order $k$.

It is not necessary to go so far. In this particular model we will  
take a short cut. The integral given by Eq.(\ref{21}) 
can be solved exactly for  $\mbox{Re}\,\,g_{0} \geq 0$, yielding
\begin{equation}
z_{0}(m_{0},g_{0})=(\frac{3}{2g_{0}})^{\frac{3}{4}}m_{0}^{2}
\Psi\left(\frac{3}{4},\frac{3}{2};\frac{3m_{0}^{4}}{2g_{0}}\right),
\label{24}
\end{equation}
where $\Psi(a,c\,;z)$ is the confluent 
hypergeometric function of second kind \cite{grads} \cite{baker}, 
and we are using the principal branch of this function.
Since we are interested in studying the analytical structure 
of $z_{0}(m_{0},g_{0})$ in the  
coupling constant complex plane at $g_{0}=0$, we must 
investigate the analytic structure of the $\Psi(a,c\,;z)$ at 
$z=\infty$.
We are following part of the discussion developed in Ref. \cite{eu}. 
The confluent hypergeometric function of second kind $\Psi(a,c\,;z)$
is a many valued analytic function of $z$, with a usual branch cut for
$|\arg\, z|=\pi$, and a singularity at $z=0$. Therefore 
$z_{0}(m_{0},g_{0})$ can be defined as a multivalued analytic function 
on the complex $g_{0}$ plane, with a branch cut for 
$|\arg\, g_{0}|=\pi$ and a singularity at $g_{0}=0$.
So we have to consider its principal branch 
in the plane cut along 
the negative real axis. The analytic continuation corresponds 
to the definition for
$z_{0}(m_{0},g_{0})$ in the whole coupling constant 
complex plane except for a branch cut for 
$|\arg\,z|=\pi$.


To generate the Schwinger functions we introduce sources in the model.
Thus we have that the zero-dimensional 
generating function  $z(m_{0},g_{0};h)$ is given by 
\begin{equation}
z(m_{0},g_{0};h)=\sqrt{\frac{2}{\pi}}
\int_{0}^
{\infty}d\varphi\,\exp\left(-\frac{1}{2}m_{0}^{2}\,
\varphi^{2}-\frac{g_{0}}{4!}\varphi^{4}\right)\cosh (h\varphi).
\label{q15}
\end{equation}
As we discussed, it is possible to find $z(m_{0},g_{0};h)|_{h=0}$ 
in a closed form. Nevertheless it is not possible to express 
$z(m_{0},g_{0};h)$ in terms of known functions. If we 
try to expand $\exp(h\varphi)$ in power 
series and, in order to solve the resulting
integrals, we interchange the summation and 
the integration, we have problems because the power series is uniformly 
convergent only if $|h\varphi|<1$. 
The result of this operation is $z^{(1)}(m_{0},g_{0};h)$ which 
is  
the asymptotic expansion of 
$z(m_{0},g_{0};h)$. We may write 
$z^{(1)}(m_{0},g_{0};h)\sim z(m_{0},g_{0};h)$. Thus we have
\begin{equation}
z^{(1)}(m_{0},g_{0};h)=\sum_{k=0}^{\infty}\,h^{2k}f_{k}(m_{0},g_{0}),
\label{q16}
\end{equation}
where the coefficients $f_{k}$ are given by
\begin{equation}
f_{k}(m_{0},g_{0})=\frac{(-1)^{k}}{\sqrt{2\pi}}\frac{2^{k+1}}{2k!}
(\frac{\partial}{\partial m_{0}^{2}})^{k}z_{0}(m_{0},g_{0}).
\label{q17}
\end{equation}
Recall that $z_{0}(m_{0},g_{0})$ is the generating function in the 
absence of sources. Using Eq.(\ref{24}) we evaluate the partial 
derivatives of $z_{0}(m_{0},g_{0})$ in the above formula.
After some algebra, we have  
the asymptotic representation for 
$z(m_{0},g_{0};h)$ in terms of derivatives of the 
confluent hypergeometric function of second kind,  
\begin{equation}
z^{(1)}(m_{0},g_{0};\,h)=
(\frac{3}{2g_{0}})^{\frac{3}{4}}\left(
\sqrt{\frac{2}{\pi}}\,m^{2}_{0}
\Psi\left(\frac{3}{4},\frac{3}{2};\frac{3m_{0}^{4}}{2g_{0}}\right)+
\sum_{k=1}^{\infty}
h^{2k}c_{k}(\frac{\partial}{\partial m_{0}^{2}})^{k}
\Psi\left(\frac{3}{4},\frac{3}{2};\frac{3m_{0}^{4}}{2g_{0}}\right)\right)
\label{qqq}
\end{equation}
where the coefficients $c_{k}$ are given by $c_{k}=
\frac{(-1)^{k}}{\sqrt{2\pi}}\frac{2^{k+1}}{2k!}$.
Let us study the singularities of $z^{(1)}(m_{0},g_{0};h)$ in the complex 
coupling constant for $0<|g_{0}|<\infty $. 
The derivatives of the confluent 
hypergeometric functions of second kind are given by 
\begin{equation}
\frac{d^{n}}{dz^{n}}\Psi(\alpha,\gamma;z)=(-1)^{n}
(\alpha)_{n}\Psi(\alpha+n,\gamma+n;z),
\end{equation}
where the coefficients $(\alpha)_{k}$ are defined by
\begin{equation}
(\alpha)_{0}=1,...\,\,(\alpha)_{k}=\frac
{\Gamma(\alpha+k)}{\Gamma(\alpha)}=\alpha(\alpha+1)...(\alpha+k-1),
\end{equation}
for $k=1,2,...$.
Therefore, again we note that in the series representation for $z^{(1)}
(m_{0},g_{0};\,h)$ we find branch points at $g_{0}=0$, 
 $g_{0}=\infty $ and a branch cut at   
$\arg\, (g_{0})=\pi $.

An interesting question is related to the zeroes of  
the confluent hypergeometric function of second kind  $\Psi(a,c\,;z)$.
Since the zero-dimensional partition function  $z_{0}(m_{0},g_{0})$ 
is proportional to $\Psi(a,c\,;z)$, $(a=\frac{3}{4}, c=\frac{3}{2})$
it is important to know if there is a ratio between $m_{0}$ and $g_{0}$
where $z_{0}(m_{0},g_{0})$ vanishes. For $a$ and $c$ real it is well 
known that  $\Psi(a,c\,;z)$ has only a finite number of positive 
zeros, and there are no zeroes for sufficiently large $z$ \cite{bateman}.
Also the confluent hypergeometric function of second kind can not have 
positive zeros if $a$ and $c$ are real and either $a>0$ or $a-1+c>0$.
Consequently, there are no real values for  $m_{0}$ and $g_{0}$ where 
the free-energy per unit volume diverges. Note that 
$\Psi(a,c\,;z)$ has complex 
zeros for real $a$ and $c$, but we are not interested in such situation. 
In the next section we repeat our zero-dimensional analysis for the 
ultra-local quantum electrodynamics.

\section{The strong-coupling expansion and the ultra-local 
approximation in quantum electrodynamics}

It is not difficult 
to study quantum electrodynamics also in the 
context of the strong-coupling expansion. We assume that such a theory 
defined in a Minkowski spacetime can be extended to the 
Euclidean formulation. Accepting this point, 
let us investigate the strong-coupling regime of the theory. The treatment is 
standard and the idea is the same as for the scalar theory, 
treating the kinetic terms of the theory as a perturbation and solving the 
self-interacting part exactly. The only point that we have to call the attention of the 
reader is that we have to introduce a mass term for the photon field.  
Here we are following the approach developed by Cooper and Kenway  \cite{co} and 
Itzykson, Parisi and Zuber \cite{ula}.  

The generating functional of all Schwinger functions for  
quantum electrodynamics defined in a d-dimensional Euclidean space is given by
\begin{equation}
Z(\bar{\eta},\eta,J_{\mu})
=\int\, d\bar{\psi}\,d{\psi}\,dA_{\mu}\,\exp\left(-\int\, 
d^{d}x\,{\cal{L}}(A_{\mu},\psi,\bar{\psi})
+sources \right),
\label{q1}
\end{equation}
where $A_{\mu}(x)$ and $\psi(x)$ are respectively the gauge and fermion field and 
${\cal{L}}(A_{\mu},\psi,\bar{\psi})$ is given by
\begin{equation}
{\cal{L}}=\frac{1}{4}(F_{\mu\nu})^{2}+M^{2}\,A_{\mu}^{2}+
\bar{\psi}(\gamma_{\mu}\partial_{\mu}+im+ie\gamma_{\mu}
A_{\mu})\bar{\psi}.
\label{q2}
\end{equation}
Note that it is necessary to introduce 
a photon mass in the same way that we did for the case of the 
scalar theory in order to regulate the photon 
part of the strong-coupling expansion. Although the 
introduction of the mass term eliminates the necessity of a gauge fixing term, 
it is useful to work in a general gauge.

The Euclidean form of the generating 
functional of complete Schwinger functions in a d-dimensional 
space in a covariant gauge is given by
\begin{equation}
Z(\bar{\eta},\eta,J_{\mu})=K_{A}\,K_{\psi}\int\, d\bar{\psi}\,d{\psi}\,dA_{\mu}\,
\exp\left(-\int\,d^{d}x\,\left(\frac{1}{2}M^{2}A^{2}+
\bar{\psi}(im+ie\gamma_{\mu}A_{\mu})\psi
+souces\right) \right)
\label{qqa}
\end{equation}
where 
\begin{equation}
K_{A}=\exp\left(\frac{1}{2}\int\,d^{d}x\,\int\, d^{d}y \frac{\delta}{\delta J_{\mu}(x)}
D^{-1}_{\mu\nu}(x,y) \frac{\delta}{\delta J_{\mu}(y)}\right),
\label{q4}
\end{equation}
and also 
\begin{equation}
K_{\psi}=\exp\left(-\int\,d^{d}x\,\int\, d^{d}y \frac{\delta}{\delta\eta(x)}S^{-1}(x,y)
\frac{\delta}{\delta\bar{\eta}(y)}\right),
\label{q5}
\end{equation}
where $\eta$ and $\bar{\eta}$ are anticommutating Grassmann sources.

In a covariant gauge we have $D^{-1}_{\mu\nu}(\alpha;x-y)=\left(\delta_{\mu\nu}\Delta-
(1-\frac{1}{\alpha})\partial_{\mu}\partial_{\nu}\right)\delta^{d}(x-y)$ and 
also $S^{-1}(x,y)=\gamma_{\mu}
\partial_{\mu}\delta^{d}(x-y)$. 
As we have been discussing, the generating functional defined by the  
remaining functional integral is a product of one-dimensional integrals in
each point of the Euclidean space. Consequently, let us study the zero-dimensional 
generating function of quantum electrodynamics defined by $z(\eta,\bar{\eta}, j)$, 
where we are choosing $m=M^{2}=1$ \cite{qed}. 
Thus in each point of the Euclidean space we have 
the following generating function:
\begin{equation}
z(\eta,\bar{\eta}, j)=\frac{1}{\pi}\int\, dA\, d\psi\,d\bar{\psi}
\exp\left(-\frac{A^{2}}{2}-\bar{\psi}(1-eA)\psi+\bar{\psi}\eta+\bar{\eta}\psi+jA\right),
\label{30}
\end{equation}
where $\bar{\psi}$ and $\psi$ are complex conjugate c-numbers variables.
Integrating over the $\bar{\psi}$ and $\psi$ variables 
it is easy to show that the generating function given by Eq.(\ref{30}) can be written as
\begin{equation}
z(\eta,\bar{\eta}, j)
=\int\,\frac{dA}{(1-eA)}\exp\left(-\frac{A^{2}}{2}+\bar{\eta}\eta(1-eA)+jA\right).
\label{31}
\end{equation}
Let us first study the generating functional $z(\eta,\bar{\eta},j)$
before using Furry's theorem and let us call this generating function at 
zero external sources as $I(v)$, were $v=-\frac{1}{e}$. Thus we have 
\begin{equation}
I(v)=v\,\int_{0}^{\infty} \frac{da}{(a+v)}\exp(-\frac{a^{2}}{2}).
\label{33}
\end{equation}
Following Stieljes \cite{knop} it 
is possible to show that if a function $F(x)$ is defined by
\begin{equation}
F(x)=\int_{0}^{\infty}\frac{f(u)}{(x+u)}du,
\label{sti}
\end{equation}
the following series $\sum_{n=1}^{\infty}a_{n}\,x^{-n}$ is an asymptotic expansion for the
integral in which the coefficients $a_{n}$ are given by
$(-1)^{n-1}a_{n}=\int_{0}^{\infty}f(u)\,u^{n-1}du,$ where $n=1,2,..$.

This generating function for zero external sources $I(z)$ 
possesses an asymptoptic expansion
\begin{equation}
I(v)\,\sim \sum_{n=1} a_{n}\,v^{-n+1},
\label{34}
\end{equation}
where the coefficients are given by $(-1)^{n-1}\,a_{n}=\int_{0}^{\infty}
\, dx\, x^{n-1}\,\exp(-\frac{1}{2}x^{2})$.
Using the fact that quantum electrodynamics is charge 
conjugate invariant, and making use of the 
Furry's theorem, we have that the generating function 
of quantum electrodynamics must be given by
\begin{equation}
z(\eta,\bar{\eta}, j)
=\int\, \frac{dA}{(1-e^{2}A^{2})^{\frac{1}{2}}}\exp\left(-\frac{A^{2}}{2}+
\bar{\eta}\eta(1-eA)^{-1}+jA\right).
\label{n1}
\end{equation}
Note that the photon propagator is given by $G=<A^{2}>$ 
and the electron propagator is $S=<\frac{1}{1-e^{2}A^{2}}>$ where 
the average is over the measure $\frac{dA}{(1-e^{2}A^{2})^{\frac{1}{2}}}\exp(-\frac{A^{2}}{2})$.

Let us proceed in the  
study of the generating function for the zero-dimensional quantum electrodynamics with 
$j$, $\bar{\eta}$ and $\eta$ being the c-number sources. In order 
to obtain some information about the structure of the singularities of the 
generating function, and also the generating functional of the Schwinger 
functions in the complex coupling constant plane, let first try to solve the integral 
defined by Eq.(\ref{n1}). It is clear that the integral that defines 
$z(\eta,\bar{\eta}, j)$ is meaningful only for $e^{2}$ negative. 
To simplify our discussion
let us first assume that 
the c-number sources $\bar{\eta}$ and $\eta$ are zero i.e. $\bar{\eta}=\eta=0$. In this 
particular situation we have
\begin{equation}
z(\eta,\bar{\eta}, j)|_{\eta=\bar{\eta}=0}=
\int\, \frac{dA}{(1-e^{2}A^{2})^{\frac{1}{2}}}\exp\left(-\frac{A^{2}}{2}+jA\right).
\label{n2}
\end{equation}
The integrand of Eq.(\ref{n2}) has two branch points at $A=\frac{1}{e}$ and $A=-\frac{1}{e}$. 
To perform the integral, let us first make the replacement
$e\rightarrow ie$. In this case the branch points appear in the imaginary axis 
of the $A$ complex plane and after we impose that our function $z_{0}(ie;\,j)$ is defined only 
for $A>0$ we have

\begin{equation}
z_{0}(ie;\,j)=
\int_{0}^{\infty}\, \frac{dA}{(1+e^{2}A^{2})^{\frac{1}{2}}}\exp\left(-\frac{A^{2}}{2}+jA\right).
\label{n3}
\end{equation}
Even after this ``improvement'', it is very difficult to express the integral 
in terms of known functions. Nevertheless an asymptoptic expansion for small $j$ 
can be found. Using a Taylor expansion for $z_{0}(ie;\,j)$ near $j=0$ we have
\begin{equation}
z_{0}(ie;\,j)=z_{0}(ie;\,j)|_{j=0}+j\frac{\partial\,z_{0}(ie;\,j)}{\partial\,j}|_{j=0}+...
\label{n5}
\end{equation}
Let us calculate first $z_{0}(ie;\,j)|_{j=0}$. We have
\begin{equation}
z_{0}(ie;\,j)|_{j=0}=
\int\, \frac{dA}{(1+e^{2}A^{2})^{\frac{1}{2}}}\exp(-\frac{A^{2}}{2}).
\label{nn3}
\end{equation}
Thus, it is not difficult to show that $z_{0}(ie;\,j)|_{j=0}$ is given by
\begin{equation}
z_{0}(ie;\,j)|_{j=0}=\frac{1}{2e}\exp\left(\frac{1}
{4e^{2}}\right)K_{0}\left(\frac{1}{4e^{2}}\right),
\label{nn4}
\end{equation}
where $K_{0}(z)$ is the Macdonald function of zero order.
It is not difficult to perform the integral that appears 
in the second term of the Taylor expansion.
A simple calculation gives

\begin{equation}
\frac{\partial z_{0}(ie;\,j)}{\partial\,j}|_{j=0}=
{\sqrt{\frac{\pi}{2e^{2}}}}\exp\left(\frac{1}{2e^{2}}
\right)\left(1-\Phi\left({\sqrt{\frac{1}{2e^{2}}}}\right)\right)
\label{qaq}
\end{equation}
where $\Phi(x)$ is the Fresnel integral defined by
\begin{equation}
\Phi(x)=\frac{2}{\sqrt{\pi}}\int_{0}^{x}\,dt\,\exp(-t^{2}).
\label{n6}
\end{equation} 
Using Eq.(\ref{n5}), Eq.(\ref{nn4}) and Eq.(\ref{qaq}) we have
\begin{equation}
z_{0}(ie;\,j)=\frac{1}{2e}\left(\exp\left(\frac{1}{4e^{2}}
\right)K_{0}\left(\frac{1}{4e^{2}}\right)+j
{\sqrt{\frac{\pi}{2e^{2}}}}\exp\left(\frac{1}{2e^{2}}\right)
\left(1-\Phi\left({\sqrt{\frac{1}{2e^{2}}}}\right)\right)\right)+..
\label{iso}
\end{equation}
It is clear that there is a 
branch point singularity in the 
complex coupling constant plane in the 
zero-dimensional model.  


In the next section we will analyse more general scalar models   
with polynomial and non-polynomial interactions in the context of the strong-coupling expansion. 
Now we turn to the second question that we have raised: 
in which circumstances the strong-coupling expansion must be used? 
To clarify this problem, in the next section we  perform an 
perturbative expansion in two toy-models where the use of the 
strong-coupling expansion is imperative.

\section{The ultra-local approximation in scalar models.}\

In the infinite cut-off limit of an infrared free theory, 
the usual perturbative expansion where we assume that the non-Gaussian
contribution is a perturbation of the corresponding free theory 
can not be used, and we resort 
to an alternative perturbative expansion. 
The strong-coupling expansion is an alternative 
perturbative expansion suitable for treating
this situation. It is remarkable that in the 
strong coupling expansion different scalar theories 
can be treated in the same way, 
since we 
factor out the free part of the Lagrangian density and evaluate the 
remaining non-Gaussian contribution in a closed form. From this discussion we see 
that this unusual expansion can be performed for any polynomial 
or non-polynomial interaction 
$V(g_{i};\varphi)$, where $g_{i},i=1,2,..n $ are the coupling constants of the model. 
In a different context, for the study of non-polynomial scalar models 
at finite temperature in the one-loop approximation, see for instance 
Ref. \cite{nfs}.

Going back, the formal representation for the 
generating functional of complete 
Schwinger functions $Z(h)$ using the 
strong-coupling expansion is given by
\begin{equation}
Z(h)=\left(1-\frac{1}{2}\int d^{d}x\,\int d^{d}y\frac{\delta}{\delta h(x)}K(m_{0},\sigma;x-y)
\frac{\delta}{\delta h(y)}+...\right)\,Q_{0}(\sigma;h),
\label{27}
\end{equation}
where the ultra-local generating functional $Q_{0}(\sigma;h)$ is 
defined by the following functional integral:
\begin{equation}
Q_{0}(\sigma;h)={\cal{N}}\int [d\varphi]\,\exp\left(\int d^{d}x\,(-\frac{1}{2}
\sigma\,m_{0}^{2}\,\varphi^{2}(x)
-V(g_{i};\varphi)+
h(x)\varphi(x))\right),
\label{28}
\end{equation}
and ${\cal{N}}$ is the normalization factor. 
Let us study the ultra-local generating functional  
$Q_{0}(\sigma;h)$ in detail. 
Note that a naive use of a continuum limit of the 
lattice regularization for the ultra-local functional integral leads
to a Gaussian theory, where we simply make use of the central 
limit theorem, which states that a probability distribution 
of the sum of $n$ independent random variables becomes Gaussian
in the limit $n\rightarrow\infty$. As ha been discussed by 
Klauder, in the limit where the coupling constant goes to 
zero, all of the interacting theory solutions become the 
pseudo-free ones. Then following Klauder we are able to represent $Q_{0}(\sigma;h)$ 
as
\begin{equation}
Q_{0}(\sigma;h)=\exp\left(-\int d^{d}x\, L(\sigma;h(x))\right),
\label{qq}
\end{equation}
where  $ L(\sigma;h(x))$ is some functional.
The formulae given by Eq.(\ref{28}) and Eq.(\ref{qq}) are 
fundamental for our study. To proceed,
let us see how it is possible to extract some informations selecting 
particular potentials $V(g_{i};\varphi)$. We limit ourselves to models 
with only one component. The generalization of our
investigations to models with more than one component 
does not present any difficulty. 
For a discussion of the strong-coupling expansion in the $O(N)$ model,
see for instance \cite{parga}.

The ideas of the preceding sections will be ilustrated in two 
models where the usual perturbative expansion in the coupling constant can not be performed. The first one is given by interaction Lagrangian 
${\cal{L}_{II}}(\beta,\gamma;\varphi)$ defined by
\begin{equation}
{\cal{L}_{II}}(\beta,\gamma;\varphi)=\beta\varphi^
{\,p}(x)+\gamma\varphi^{-p}(x).
\label{q12}
\end{equation}
where $\beta$ and $\gamma$ are bare parameters and $p$ is an integer.
The second model that we would like to discuss, which we call the 
sinh-Gordon model, is defined by
the following interaction Lagrangian:
\begin{equation}
{\cal{L}_{III}}
(\beta,\gamma;\varphi)=\beta\left(\cosh \gamma\varphi(x)-1\right),
\label{q13}
\end{equation}
where $\beta$ and $\gamma$ are also bare parameters.
Let us start studying the model defined by Eq.(\ref{q12}).
Note that in this model we have a suppression of the 
configurations fluctuations around $\varphi=0$.
For $p$ even the model has two minima. It is not difficult to show that the 
interaction Lagrangian density of the
model has a 
power series representation. The equilibrium values are given by $\varphi_{0}=
(\frac{\gamma}{\beta})^{\frac{1}{2p}}$ and $\varphi_{0}=-
(\frac{\gamma}{\beta})^{\frac{1}{2p}}$. Let us choose the case where 
$\varphi_{0}>0$ and define a new field $\phi(x)=(\varphi(x)-\varphi_{0})$. Using the 
binomial expansion and its generalization
\begin{equation}
(1+x)^{\alpha}=\sum_{k=0}^{\infty}\,
\left(_{k}^{\alpha}\right)\,x^{k},\,\,\,|x|<1,
\label{48}
\end{equation}
where the coefficients of the expansion are given by
\begin{equation}
\left(^{\alpha}_{0}\right)=1,\,\,\,
\left(^{\alpha}_{k}\right)=\frac{\alpha(\alpha-1)...(\alpha-k+1)}{k!},\,\,
\,for \,\, k\geq 1,
\end{equation} 
we have 
\begin{equation}
{\cal{L}_{II}}(\gamma,\beta;\,\phi)=\sum_{k=0}^{p}c(p,k)\,\phi(x)^{k}+
\gamma\,\varphi_{0}^{-p-k}\sum_{k=p+1}^{\infty}
\left(_{k}^{-p}\right)\,\phi(x)^{k}.
\end{equation}
The coefficients $c(p,k)$ are given by
\begin{equation}
c(p,k)=\left(\beta\,\varphi_{0}^{-p-k}\left(_{k}^{p}\right)+\gamma\,
\varphi_{0}^{-p-k}\left(_{k}^{-p}\right)\right).
\end{equation}
Note that the generalization of the binomial series is valid for 
any complex exponent $\alpha$. In other words, 
the power series in Eq.(\ref{48}) is convergent everywhere in the $\alpha$
complex plane.
Since we are interested in the non-perturbative effects, let us study the 
ultra-local version of the model.
The zero-dimensional generating functional is given by 
\begin{equation}
z_{2}(\beta,\gamma;\,h)=\sqrt{\frac{2}{\pi}}\int_{0}^{\infty}
d\varphi\,
\exp\left(-\beta\varphi^{\,p}-\gamma\varphi^{-p}\right)
\cosh (h\varphi).
\label{oque}
\end{equation}
It is not possible to express $z_{2}(\beta,\gamma;\,h)$ 
in terms of known functions. 
First let us express 
the zero-dimensional generating function $z_{2}(\beta,\gamma;\,h)$
in the absence of sources in a closed form.  
Then, expand $\cosh(h\varphi)$ in power 
series and, in order to solve the resulting integrals, interchange the summation and 
the integration. Again $z^{(2)}(\beta,\gamma;\,h)$ 
that we obtained after use the power series expansion is the asymptotic expansion of 
$z_{2}(\beta,\gamma;\,h)$ and we write that
$z^{(2)}(\beta,\gamma;\,h)\sim z_{2}(\beta,\gamma;\,h)$. 
It is not difficult to find $z_{2}(\beta,\gamma;\,h)|_{h=0}$.  
Using the identity
\begin{equation}
\int_{0}^{\infty}\,dx\,\,x^{\nu-1}\exp\left(-\beta\, x^{\,p}-\gamma\, x^{-p}\right)=
\frac{2}{p}(\frac{\gamma}{\beta})^{\frac{\nu}{2p}}K_{\frac{\nu}
{p}}\left(2{\sqrt{\beta\gamma}}\right),
\label{inn4}
\end{equation}
that is valid for $Re\,\beta>0$ and $Re\, \gamma>0$, 
and where $K_{\nu}(z)$ is the modified Bessel function of third kind,
we have 
that the zero-dimensional generating function $z_{2}(\beta,\gamma;\,h)$
in the absence of sources is given by
\begin{equation}
z_{2}(\beta,\gamma;\,h)|_{h=0}=\frac{1}{p}\sqrt{\frac{8}{\pi}}
(\frac{\gamma}{\beta})^{\frac{1}{2p}}
K_{\frac{1}{p}}\left(2\,{\sqrt{\beta\gamma}}\right).
\label{xuva}
\end{equation}
Using that $z^{(2)}(\beta,\gamma;\,h)\sim\, z_{2}(\beta,\gamma;\,h)$ we have 
\begin{equation}
z^{(2)}(\gamma,\beta,h)=\sum_{k=0}^{\infty}h^{2k}c(p,k)
(\frac{\gamma}{\beta})^{\frac{2k+1}{p}}K_
{\frac{2k+1}{p}}\left(2{\sqrt{\beta\gamma}}\right),
\end{equation}
where the coefficients $c(p,k)$ are  given by 
\begin{equation}
c(p,k)=\frac{1}{p}\sqrt{\frac{8}{\pi}}\frac{1}{2k!}.
\label{ultimo}
\end{equation}

Let us discuss now the sinh-Gordon model. 
This model is also non-renormalizable 
in the weak-coupling-perturbative expansion, 
where $\beta$ and $\gamma$ are the coupling constants and we choose $\sigma=0$. It is clear that the zero-dimensional generating function $z_{3}(\beta,\gamma;\,h)$
in the absence of sources can be found in a closed form and it is given by:
\begin{equation}
z_{3}(\beta,\gamma;\,h)|_{h=0}=\frac{2e^{\beta}}{\gamma}K_{0}(\beta). \label{mara}
\end{equation}
It is interesting that for this 
kind of theory, even in the presence of sources, 
we can find a closed form for the 
zero-dimensional generating fuction $z_{3}(\beta,\gamma;\,h)$,
\begin{equation}
z_{3}(\beta,\gamma;\,h)=\frac{2e^{\beta}}{\gamma}K_{\frac{h}{\gamma}}(\beta),
\label{mar}
\end{equation}
where again $K_{\nu}(z)$ is the modified Bessel function of third kind. 
Using Eq.(\ref{qq}), Eq.(\ref{mara}) and Eq.(\ref{mar}) 
we have that the ultra-local generating 
functional for the sinh-Gordon model is given by 
\begin{equation}
Q_{0}(h)={\cal{N}}
\exp\left(\delta^{d}(0)\,\int\,d^{d}x\,\ln\,\left(\frac{2e^{\beta}}
{\gamma}K_{\frac{h}{\gamma}}(\beta)\right)\right),
\label{ult}
\end{equation}
where the normalization $N$ can be found using that $Q_{0}(h)|_{h=0}=1$.
It is remarkable that in the Sinh-Gordon model it is possible to 
extract information from a finite number of terms of the infinite series 
representation of the generating functional in the strong coupling 
expansion. In the Sinh-Gordon model the ultra-local approximation is exact.
Thus $Z(h)\equiv Q_{0}(h)$. 
Let us calculate the ``free-energy'' per unit volume $f(\beta,\gamma;\,h)$ 
for $\beta\neq 0$ and also $\gamma\neq 0$.
It is clear that we have 
\begin{equation}
f_{ren}(\beta,\gamma;\,h)=\ln\left(\frac{2\,e^{\beta}}{\gamma}\right)+
\ln\left(K_{\frac{h}{\gamma}}(\beta)\right).
\end{equation}
The first question that can be asked is related to the 
fact that $f_{ren}(\beta,\gamma;\,h)$ can diverge, since 
the modified Bessel function of third kind $K_{\nu}(z)$ has zeros in the complex plane. From this discussion 
one needs information about the distribution of the zeros of the 
modified Bessel function of third kind 
in the complex plane. For real $\nu$, $K_{\nu}(z)$ 
has no zeros in the region $|arg\,z|<\frac{\pi}{2}$ and 
in the complex plane with a cut along the segment $(-\infty,0]$ 
the $K_{\nu}(z)$ has a finite number 
of zeros. Thus the free energy per unit volume in the model is finite. In the next section we sketch how it is possible to obtain a renormalized Schwinger functional going beyond the ultra-local approximation.

\section{Going beyond the ultra-local approximation in scalar models.}\

In the next section we will be interested in computing global quantities, as 
for example the free energy or the pressure of the vacuum. The 
picture that emerges from our discussion is the following:
in the strong-coupling perturbative expansion we reduce the problem of the 
singularities of the Schwinger functional into two parts. The first one 
is how to define the ultra-local generating functional and the second one is 
to regularize and renormalize the other terms of the perturbative expansion,
which came from the coupling between distinct points 
and are giving by the 
non-diagonal part of the kernel. Let us assume that a scalar field with some generic 
interaction Lagrangian is defined in a compact region of the Euclidean 
space. Thus it is clear that the zeta function regularization  
can be used to control the divergences of the kernel 
$K(m_{0};\,x-y)$ integrated over the volume \cite{hawking}.

In the infinite volume limit,
to complete our discussion, we have to 
sketch the formalism that can be used to obtain a regularized 
expression for $Z^{(1)}(h)$ going beyond the ultra-local approximation. 
We 
are using two different regularization procedures, and it is possible 
to identify the divergent contribution
in each regularized expression and a renormalization procedure 
is implemented with an appropriate 
subtraction of the singular contribution. 
We would like to stress that such kind of study 
in another context was performed by Svaiter and Svaiter \cite{ss}. 
These authors developed a method 
to unify two unrelated regularization methods frequently employed to 
obtain the renormalized
zero-point energy of quantum fields. Introducing a mixed 
cut-off function and studying the analytic properties of 
the regularized energy as a function of the two cut-off parameters it 
was possible to not only relate the usual cut-off method and 
the analytic regularization method, 
but also unify both methods.

Let us use the ideas discussed above introducing 
a exponential cut-off and also an algebraic cut-off to regularize the kernel
$K(m_{0},x-y)$. For simplicity let us assume that we have a constant external source,
i.e., $h(x)=h=constant$ and also $\sigma=0$.
The second term of the perturbative series given by Eq.(\ref{q3}) becomes 
\begin{equation}
Z^{(1)}(h)=-\frac{1}{2}\frac{\partial^{2}Q_{0}(h)
}{\partial\,h^{2}}
\int d^{d}x\,\int d^{d}y\,K(m_{0};\,x-y).
\label{pu1}
\end{equation}
The singularities of the $Z^{(1)}(h)$ are coming from different terms. 
First the second derivative with respect to the source of the ultra-local 
generating functional is singular in the continuum limt. Since we have been 
discussing how to deal with the singularities of the ultra-local 
generating functional, let us study only the divergences coming from
the kernel $K(m_{0},x-y)$ integrated over the Euclidean 
volume where the scalar field has been defined.
In the infinite volume limit, the Fourier 
representation for the kernel $K(m_{0};\,x-y)$ is given by 
\begin{equation}
K(m_{0};\,x-y)=\frac{1}{(2\pi)^{d}}\int\,d^{d}q\,
(q^{2}+m_{0}^{2})\,\exp\left(iq(x-y)\right).
\label{pu3}
\end{equation}
To evaluate the behavior of the kernel $K(m_{0};\,x-y)$ for $|x-y|$ small and large, 
as we discussed let us introduce two different regulators.  
The divergent expression given by Eq.(\ref{pu3}) can 
be regularized using for example a exponential cut-off 
function $f_{1}(m_{0},\eta;\,q)$ defined by
\begin{equation}
f_{1}(m_{0},\eta;\,q)=
\exp\left(-\eta\,(q^{2}+m_{0}^{2})\right),\,\,\,\,\mbox{Re}(\eta)>0.
\label{pu4}
\end{equation}
Another possibility is to use an analytic regularization procedure introducing an
algebraic cut-off function $f_{2}(m_{0}.\rho;\,q)$ defined by 
\begin{equation}
f_{2}(m_{0},\rho;\,q)=
(q^{2}+m_{0}^{2})^{\rho},\,\,\,\,\mbox{Re}(\rho)<-\frac{d}{2}-1.
\label{pu5}
\end{equation}
It is clear that the analytic regularization that we are using is similar to 
the analytic and dimensional regularization used to control divergences of the 
Feynman diagrams in the weak coupling expansion \cite{dimensional}.
Explicit and exact 
integrations can be performed in both cases.

In order to carry out this program, let us study first the exponential 
cut-off method. The regularized kernel $K(m_{0},\eta;\,x-y)$ is defined by
\begin{equation}
K(m_{0},\eta;\,x-y)=\frac{1}{(2\pi)^{d}}\int\,d^{d}q\,
\exp\left(iq(x-y)\right)\,(q^{2}+m_{0}^{2})
\exp\left(-\eta(q^{2}+m_{0}^{2})\right).
\label{pu6}
\end{equation}
It is clear that we can write the regularized kernel $K(m_{0},\eta;\,x-y)$ as
\begin{equation}
K(m_{0},\eta;\,x-y)=-\frac{1}{(2\pi)^{d}}\frac{\partial}
{\partial\eta}\int\,d^{d}q\,
\exp(iq(x-y))\,\exp\left(-\eta(q^{2}+m_{0}^{2})\right).
\label{pu7}
\end{equation}
Since the integral in Eq.(\ref{pu7}) is Gaussian it can be 
performed, and we obtain the following expression for the regularized kernel:
\begin{equation}
K(m_{0},\eta;\,x-y)=-\frac{1}{(2{\sqrt{\pi}})^{d}}
\frac{\partial}{\partial\eta}\left(\eta^{-\frac{d}{2}}
\exp(-\eta\,m_{0}^{2}-\frac{1}{4\eta}
(x-y)^{2})\right).
\label{pu8}
\end{equation}
Thus we have that the regularized kernel 
$K(m_{0},\eta;\,x-y)$ can be expressed as
\begin{eqnarray}
&&K(m_{0},\eta;\,x-y)=\nonumber \\
&&-\frac{1}{(2{\sqrt{\pi}})^{d}}\,
\exp\left(
-\eta\,m_{0}^{2}
-\frac{1}{4\eta}
(x-y)^{2}\right)\left(-\frac{1}{4\eta^
{\frac{d}{2}+2}}(x-y)^{2}+\frac{d}{2\eta^{\frac{d}{2}+1}}
+\frac{m^{2}_{0}}{\eta^{\frac{d}{2}}}\right).
\label{pu9}
\end{eqnarray}
Note that the negative powers portion of the Laurent series expansion of 
$K(m_{0},\eta;\,x-y)$ around $\eta=0$ has an infinite number of terms and the 
regularized expression has an essential singularity at $\sigma=0$.
Thus, let use an alternative method, i.e. 
the analytic regularization procedure, 
that we call an algebraic cut-off. The same idea was presented by Kovesi-Domokos \cite{kovesi}
in the regularization of the strong-coupling perturbative expansion. 

Using the algebric cut-off function $f_{2}(m_{0},\rho;\,q)$, the regularized kernel 
$K(m_{0},\rho;\,x-y)$ is defined by
\begin{equation}
K(m_{0},\rho;\,x-y)=\frac{1}{(2\pi)^{d}}
\int\,d^{d}q\,\exp\left(iq(x-y)\right)\,(q^{2}+m_{0}^{2})^{1+\rho}.
\label{pu10}
\end{equation}
The regularized kernel $K(m_{0},\rho;\,x-y)$ is 
convergent and analytic in the 
complex $\rho$ plane for $\mbox{Re}(\rho)<-\frac{d}{2}-1$. As 
in any cut-off method we have 
to take the limit $\rho\rightarrow\, 0$, 
starting from $\mbox{Re}(\rho)<-\frac{d}{2}-1$. To 
perform the d-dimensional integration let us work in a d-dimensional polar coordinate 
system. Defining $|x-y|=r$ and $q=(q_{1}^{2}+q_{2}^{2}+..+q_{d}^{2}\,)^
{\frac{1}{2}}$ it is easy to show that
the regularized kernel can be expressed in the following way:
\begin{equation}
K(m_{0},\rho;\,r)=\frac{1}{(2\pi)^{d}
r^{\frac{d}{2}-1}}\,\int_{0}^{\infty}
\,dq\, q^{\frac{d}{2}}(q^{2}+m_{0}^{2})^{1+\rho}
J_{\frac{d}{2}-1}(q\,r),
\label{pu11}
\end{equation}
where $J_{\nu}(z)$ is the Bessel function of first kind 
of order $\nu$. Let us analyse the cases $r\neq 0$ and the 
case where $r=0$ separately to make our discussion more 
precise:

i) the case $r=0$: this one is trivial. 
We have for odd $d$ that the kernel is given by 
$K(m_{0},\rho;\,x\approx y)|_{\rho=0}=0$ and 
for even $d$ it is trivial to show that
\begin{equation}
K(m_{0},\rho;\,x\approx y)=\frac{1}
{(2{\sqrt{\pi}})^{d}}(m_{0}^{2})^{\frac{d}{2}+\rho+1}.
\label{pu14}
\end{equation}

ii) the case $r\neq 0$:
the more interesting case, where $r\neq 0$ can be solved 
evaluating the integral $I(\mu,\nu;\,a,b)$ defined by
\begin{equation}
I(\mu,\nu;\,a,b)=
\int_{0}^{\infty}\,dx\,\frac{x^{\nu+1}}{
(x^{2}+a^{2})^{\mu+1}}J_{\nu}(bx),
\,\,\,\, a>0,\,\,b>0.
\label{pu15}
\end{equation}
To evaluate this integral first let us start from an 
integral representation for the gamma function $\Gamma(z)$, using the identity
\begin{equation}
\frac{1}{(x^{2}+a^{2})^{\mu+1}}=
\frac{1}{\Gamma(\mu+1)}\int_{0}^{\infty}\,dt\,t^{\mu}
\exp\,\left(-t(x^{2}+a^{2})\right),
\label{pu16}
\end{equation}
which is valid for $\mbox{Re}(\mu)>-1$
and, in order to have absolute convergence of the double integral 
that we are evaluating, we assume that 
the parameters $\mu$ and $\nu$ are defined in the region 
$-1<\mbox{Re}(\nu)<2\,\mbox{Re}(\mu)+\frac{1}{2}$. Then, using an integral representation for the 
Macdonald's function and also the identity given by
\begin{equation}
\int_{0}^{\infty}\,dx\, x^{\nu+1}
\exp(-a^{2}x^{2})\,J_{\nu}(bx)=
\frac{b^{\nu}}{(2a^{2})^{\nu+1}}\exp\left(-\frac{b^{2}}{4a^{2}}\right),\,\,\, a>0,\,\,\,b>0,   
\label{pu17}
\end{equation}
it is possible to show that $I(\mu,\nu;\,a,b)$ is given by
\begin{equation}
I(\mu,\nu;\,a,b)=\frac{a^{\nu-\mu}b^{\mu}}{2^{\mu}\Gamma(\mu+1)}K_{\nu-\mu}(ab),
\label{pu12}
\end{equation}
which is valid for $-1<\mbox{Re}(\nu)<2\,\mbox{Re}(\mu)+\frac{1}{2}$. Using 
the principle of analytic 
continuation we have that the regularized kernel is given by
\begin{equation}
K(m_{0},\rho;\,r)=\frac{1}{(2\pi)^{d}}\,
\left(\frac{m_{0}}{r}\right)^{\frac{d}{2}+\rho+1}
\Gamma(-1-\rho)^{-1}
K_{\frac{d}{2}+\rho+1}(m_{0}\,r).
\label{pu13}
\end{equation}
We obtained that in the limit where $\rho\rightarrow 
0$ only the $r=0$ case gives  
contribution to 
$Z^{(1)}(h)$, since the Gamma function $\Gamma(z)$ has 
simple poles at $z=0,-1,-2...$ 
and $\frac{1}{\Gamma(z)}$ is an entire function of $z$. 
From our discussions we see that the divergence that appears in the 
strong-coupling perturbative expansion in the first-order approximation 
is proportional to the 
volume of the domain where we defined the fields.

\section{The renormalized vacuum energy beyond the one-loop level}

Finally, in this section we would like to show that
using the strong-coupling perturbative expansion 
it is possible to compute the renormalized vacuum energy 
of a self-interacting scalar field going beyong the one-loop level.
Actually our results are between the one-loop and the two-loop results.

The Casimir energy, or the 
renormalized vacuum energy of a free quantum field in the 
presence of macroscopic boundaries, can be derived using 
the formalism of the path integral quantization and 
the zeta-function regularization \cite{hawking} or a different method
as the Green's function method. In fact, these results are at one-loop level. Although, higher-order loop corrections seems now beyond the experimental reach, at least theoretically such corrections are of interest, as has 
been stressed by Milton \cite{milton}. 

In the present chapter we  
analyse the important question of the 
perturbative expansion of a self-interacting scalar field in the
presence of boundaries that break translational invariance.
Using the strong-coupling perturbative expansion 
we will show how it is possible to compute the renormalized vacuum energy 
of a self-interacting scalar field going beyong the one-loop level.

Starting from the Schwinger functional it is possible to define 
the generating functional of connected correlation functions, given 
by $\ln Z(h)$. Consequently we have 
\begin{equation}
\ln Z(h)=\ln\left(Q_{0}(h)-\frac{1}{2}\frac{\partial^{2}Q_{0}(h)
}{\partial\,h^{2}}
\int d^{d}x\,\int d^{d}y\,K(m_{0};\,x-y)\,\,+...\right),
\label{uuu1}
\end{equation}
where we are assuming that the external source $h(x)$ is constant. 
At the first-order approximation we have
\begin{equation}
\ln Z(h)=-\sum_{k=1}^{\infty}\frac{1}{2k}\left(
\frac{1}{Q_{0}(h)}\frac{\partial^{2}Q_{0}(h)}{\partial\,h^{2}}
\int d^{d}x\,\int d^{d}y\,K(m_{0};\,x-y)\,\right)^{k}.
\label{uuu2}
\end{equation}
We will concentrate only in the leading-order approximation, yielding
for the generating functional of connected correlation functions

\begin{equation}
\ln Z(h)=-\frac{1}{2Q_{0}(h)}\frac{\partial^{2}Q_{0}(h)}{\partial\,h^{2}
}\int d^{d}x\,\int d^{d}y\,K(m_{0};\,x-y).
\label{uuu3}
\end{equation}
Using the results of the last section we can write the free-energy 
per unit volume $\frac{\ln Z(h)}{V}$ as 
\begin{equation}
\frac{\ln Z(h)}{V}=-\frac{1}{2Q_{0}(h)}\frac{\partial^{2}Q_{0}(h)}{\partial\,h^{2}}
\,f(m_{0},g_{0}),
\label{uuu4}
\end{equation}
where $f(m_{0},g_{0})$ is an analytic function of $m_{0}$ and $V$ is the 
volume of the compact space. Note that we still have to investigate 
the contribution coming from the ultra-local generating functional 
$Q_{0}(h)$. 

Next let us use the cumulant expansion, which relates the average of the 
exponential to the exponential of the average. In our case we have the  
following expansion
\begin{equation}
<e^{\Omega}>=e^{\left(<\Omega>+\frac{1}{2}
(<\Omega^{2}>-<\Omega>^{2})+..\right)}
\label{uuu5}
\end{equation}
From the above equation we can see that at the first order approximation
the free-energy associated with the self-interacting field in the 
presence of boundaries is given by
\begin{equation}
\ln Z(h)=-\frac{1}{2Q_{0}(h)}\frac{\partial^{2}Q_{0}(h)
}{\partial\,h^{2}}\,\,Tr\left(e^{K(m_{0};x-y)}\right). 
\label{uuu6}
\end{equation}
The above equation is quite interesting. We have first a contribution that
is the one-loop level and using the heat-kernel expansion it is possible to compute the one-loop vacuum energy, but the ultra-local generating functional give a correction to this one-loop result. 
Consequently we are between the one-loop and the two-loop 
level.

\section{Conclusions}

 In this paper, we first discuss the usual weak-coupling perturbative 
expansion and the problems presented by it. 
The weak-coupling Feynman-Dyson perturbative expansion with the 
respective Feynman diagrams is a general method to calculate the 
Green's functions of a renormalizable model in field theory.
Since it is possible to express the $S$ matrix elements in terms of
the vacuum expectation values of products of the field operator, in 
particle physics it seems natural to perform in the majority of 
situations the weak-coupling perturbative expansion. 
Also, when studying for example the $(\lambda\varphi^{4})_{d}$  
model, in a four dimensional Euclidean space,
since the free field Gaussian model gives 
a correct description of the critical 
regime when $d>4$, it is natural to perform a perturbative 
expansion with respect to the anharmonic terms of the Lagrangian density. 
Nevertheless, as we discussed, there are many situations where the usual 
perturbative expansion can not be used. 
Consequently, we used a non-standard perturbative approach 
that have been called in the literature the strong-coupling expansion.

From the ultra-local generating functional 
in different models we obtained 
non-perturbative results.
We first analyse the singularities of 
the strong-coupling perturbative expansion in the $(\lambda\varphi^{4})_{d}$ 
model and also quantum electrodynamics 
defined in a d-dimensional Euclidean space. 
From the discussions we can see that the divergences which occur 
in any scalar model in the strong coupling expansion  
fall into two distinct classes. The first 
class is related to the 
infinite volume-continuum hypotesis for the physical 
Euclidean space. The second one is related to the 
functional form of the interaction action.
We showed that there is a branch point singularity
in the complex coupling constant plane
in the $\lambda\varphi^{4}$ model and also in quantum electrodynamics. 
Second, we discuss the ultra-local generating functional in  
two field theory toy-models
defined by the following interaction Lagrangians: 
${\cal{L}_{II}}(\beta,\gamma;\varphi)=
\beta\varphi^{\,p}(x)+\gamma\varphi^{-p}(x)$, and 
the sinh-Gordon model $({\cal{L}_{III}}(\beta,\gamma;\,\varphi)=
\beta\left(\cosh (\gamma\,\varphi)-1)\right)$. A careful analysis of the
analytic structure of both zero-dimensional partition functions in 
a four dimensional complex $(\beta,\gamma)$ space is 
still under investigation.

Performing our expansion in bounded Euclidean volume,
we sketch how it is possible to obtain a renormalized generating functional 
for all Schwinger functions, going beyond the ultra-local 
approximation. Note that the
picture that emerges from the discussion is the following: in the 
strong-coupling perturbative expansion we split the problem of defining the 
Schwinger functional in two parts: the first is how to define precisely
the ultra-local generating functional. Here is mandatory the use of a lattice appoximation to give a mathematical 
meaning to the non-Gaussian functional (actually it is not easy 
to recover the 
continuum limit).
The second part is to go beyond the ultra-local 
approximation and taking into account the perturbation part. 
This problem can be controled using 
an analytic regularization in the continuum. Besides these technical problems we still have the problem of obtaning the Schwinger functions from this approach. By summing over all the terms of the perturbative  expansion these difficulties might be solved. We believe that the strong-coupling perturbative expansion is not adequate to obtain the Schwinger functions of the models. In the strong-coupling perturbative expansion 
the free energy, or any quantity that can be derived  from the free
energy, can be obtained. An interesting application is to calculate the 
free energy associated with self-interacting fields going beyond 
the one-loop level.

The renormalized vacuum energy of free quantum fields has been derived using 
different methods, as for example the zeta-function regularization \cite{hawking}. In 
fact, the majority of the results in the literature are at the one-loop level.
Although 
higher-order loop corrections seems now beyond the experimental reach, 
at least theoretically such corrections are of interest. In the 
strong-coupling perturbative expansion we split the problem of defining the 
Schwinger functional in two parts: the ultra-local generating functional
and the perturbation part that can be controled using the heat kernel, 
zeta function regularization or any analytic regularization in the continuum. 
Consequently, a natural extension of this work in
to study the renormalized zero-point energy of interacting fields confined
in a finite volume. The strong-coupling perturbative expansion 
is an alternative method to compute the renormalized 
free energy or the vacuum energy associated 
with self-interacting fields, going beyond the one-loop level. 
This topic will be a subject of future investigation by the author.   

\section{Acknowledgements}

I would like to thank the Center of Theoretical Physics, Laboratory 
for Nuclear Science and Department of Physics, Massachusetts 
Institute of Technology,  where part of this work
was carried out, for the hospitality. I would like to acknowledge 
R.D.M. De Paola, M.Caicedo,  
S. Joffily and B. Schroer for fruitful discussions.
This paper is a slight modified version of a preliminar version that 
was published in the Proceedings of the X Brasilian School of Cosmology
and Gravitation. This paper was supported by the Conselho Nacional de
Desenvolvimento Cientifico e Tecnol{\'o}gico do Brazil (CNPq).

\end{document}